\documentclass[11pt]{article}
\usepackage[utf8]{inputenc}
\usepackage{booktabs}
\usepackage{cite}
\usepackage{amsmath,amssymb,amsfonts}
\usepackage{amssymb}
\usepackage{algorithmic}
\usepackage{graphicx}
\usepackage{textcomp}
\usepackage{xcolor}
\usepackage{xspace}
\usepackage{color}
\usepackage{tcolorbox}
\usepackage{multirow}
\usepackage{multicol}
\usepackage{booktabs}
\usepackage[caption=false]{subfig}
\usepackage{url}
\usepackage{balance}
\usepackage{xspace}
\usepackage{float}
\usepackage{setspace}
\usepackage[paperwidth=8.5in,paperheight=11in,left=1in,right=1in,top=1.5in,bottom=1.5in]{geometry}
\usepackage{picins}
\usepackage{subfig}
\usepackage{adjustbox}
\usepackage{color,soul}
\usepackage{listings}
\usepackage[hidelinks]{hyperref}
\usepackage{orcidlink}
\usepackage{listings}
\usepackage{enumitem}
\usepackage{xcolor}
\usepackage{color, colortbl}
\usepackage{subcaption}
\usepackage{float}
\usepackage{authblk}
\usepackage{graphicx}
\usepackage{float}
\usepackage{caption}

\newcommand{\tcode}[1]{{\sc #1}}

\newcommand{\lessonlearned}[1]{
\noindent \fcolorbox{black}{mygray}{\parbox{0.983\textwidth}{ #1 }} \vspace{0.5em}
}

\definecolor{mygray}{gray}{0.98}
\definecolor{lightergray}{RGB}{240,240,240}
\definecolor{lightgray}{rgb}{.9,.9,.9}
\definecolor{darkgray}{rgb}{.4,.4,.4}
\definecolor{purple}{rgb}{0.65, 0.12, 0.82}

\newfloat{tmplist}{htbp}{loc}
\floatname{tmplist}{Prompt}

\DeclareCaptionFormat{custom}
{%
    {\small \textbf{#1#2} #3}
}
\newcommand{\prompt}[3]{
    \vspace{1.5em}

    \noindent \fbox{
        \begin{minipage}{\dimexpr\textwidth-2\fboxsep-2\fboxrule\relax}
            \vspace{0.5em}
            {\small \textit{#1}}
            \vspace{0.5em}
        \end{minipage}
    }
    
    \vspace{1.5em}
}

\usepackage{todonotes}
\usepackage{soul}

\makeatletter
\def\namedlabel#1#2{
    \begingroup
    \def\@currentlabel{#2}%
    \label{#1}
   \endgroup
}
\makeatother

\lstdefinelanguage{JavaScript}{
  keywords={typeof, new, true, false, catch, function, return, null, catch, switch, var, if, in, while, do, else, case, break},
  keywordstyle=\color{blue}\bfseries,
  ndkeywords={class, export, boolean, throw, implements, import, this},
  ndkeywordstyle=\color{darkgray}\bfseries,
  identifierstyle=\color{black},
  sensitive=false,
  comment=[l]{//},
  morecomment=[s]{/*}{*/},
  commentstyle=\color{purple}\ttfamily,
  stringstyle=\color{red}\ttfamily,
  morestring=[b]',
  morestring=[b]"
}

\begin{document}
\onehalfspacing


\title{NoCodeGPT: A No-Code Interface for Building Web Apps with Language Models}

\author[1,3]{Mauricio Monteiro}
\author[2]{Bruno Castelo Branco}
\author[2]{Samuel Silvestre}
\author[2]{Guilherme Avelino}
\author[1]{Marco Tulio Valente}
\affil[1]{\normalsize Department of Computer Science, UFMG, Brazil (corresponding authors)}
\affil[2]{\normalsize Department of Computer Science, UFPI, Brazil}
\affil[3]{\normalsize Department of Industrial Automation and Information Technology, IFMG, Brazil}

\date{mauricio.monteiro@ifmg.edu.br, \{bruno.branco,sssb,gaa\}@ufpi.edu.br, mtov@dcc.ufmg.br}

\maketitle

\begin{abstract}

\noindent  In this paper, we first report an exploratory study where three participants were instructed to use ChatGPT to implement a simple Web-based application. A key finding of this study revealed that ChatGPT does not offer a user-friendly interface for building applications, even small web systems. For example, one participant with limited experience in software development was unable to complete any of the proposed user stories. Then, and as the primary contribution of this work, we decided to design, implement, and evaluate a tool that offers a customized interface for language models like GPT, specifically targeting the implementation of small web applications without writing code. This tool, called NoCodeGPT, instruments the prompts sent to the language model with useful contextual information (e.g., the files that need to be modified when the user identifies and requests a bug fix). It also saves the files generated by the language model in the correct directories. Additionally, a simple version control feature is offered, allowing users to quickly revert to a previous version of the code when the model enters a hallucination process, generating worthless results. To evaluate our tool, we invited 14 students with limited Web development experience to implement two small web applications using only prompts and NoCodeGPT. Overall, the results of this evaluation were quite satisfactory and significantly better than those of the initial study (the one using the standard ChatGPT interface). More than half of the participants (9 out of 14) successfully completed the proposed applications, while the others completed at least half of the proposed user stories.\\


\end{abstract}

\section{Introduction}
\label{sec::introduction}

Large-scale Language Models (LLM) are experiencing significant adoption among software developers, with some studies reporting major improvements in productivity. For example, a recent study by GitHub, based on telemetry data from nearly one million users, concluded that developers tend to accept 30\% of the code suggestions provided by the Copilot tool.\footnote{https://github.blog/2023-06-27-the-economic-impact-of-the-ai-powered-developer-lifecycle-and-lessons-from-github-copilot/} The study extrapolates that this adoption rate is equivalent to adding 15 million developers to the global workforce of software professionals. It also concludes that these productivity gains will have a significant impact as “developers seize new opportunities to utilize AI for solutions design and accelerate digital transformation worldwide”.

Other studies investigated the usage and benefits of language models in specific software engineering tasks, including fixing bugs~\cite{SobaniaBHP23, XiaWZ23}, writing unit tests~\cite{siddiq2023exploring}, writing code comments~\cite{geng2023large,shin2023prompt}, and solving programming problems~\cite{msr-copilot, Dakhel2023}. However, to the best of our knowledge, {\bf there are few papers that investigate the use of language models for the end-to-end construction of software systems}, i.e., in contexts where a developer has a set of requirements and has to design, implement, test, and validate an entirely new system. In such contexts, language models are used as code generators, receiving as input prompts describing the functional and non-functional requirements of a system and producing as output a runnable software application. One exception is a paper by Peng et al.~\cite{peng2023impact} in which the authors use GitHub Copilot to implement a web server starting from a high-level textual description. However, this implementation is relatively small (since it is a simple web server) and it is performed entirely in JavaScript. In other words, it does not follow widely adopted software architectures, such as an architecture organized into front-end and back-end components, which is widely used by Web-based systems nowadays. The implementation also does not use of popular frameworks for building Web interfaces or database systems. Another example refers to a demo conducted by one of the co-founders of OpenAI during the launch of GPT-4.\footnote{https://www.youtube.com/live/outcGtbnMuQ} To demonstrate the power of the new version, he drew a mockup of an application on a napkin, and from a photo of that mockup, ChatGPT was able to generate a functional web application, although apparently consisting only of its front-end component. 

In this article, we start by reporting an exploratory study using ChatGPT version 4 to implement a simple Web-based application. We defined a set of user stories and technologies for the system. Then, three developers with different profiles and experience independently attempted to implement this application using only ChatGPT. 

As a key finding of this first study, it became clear that {\bf ChatGPT does not offer a user-friendly interface for building applications, even small web systems}. In fact, the two participants with web development experience were able to build the proposed system. However, the third participant, with limited experience in software development, was unable to complete any of the proposed user stories. For example, he struggled to copy the code generated by ChatGPT into the correct folders (as defined by the system's architecture). More importantly, when ChatGPT produced an incorrect sequence of results, he was not able to revert to the last stable version that had been generated by the model.

Thus, as the primary contribution of this work, we  designed, implemented, and evaluated a {\bf tool that offers a customized interface for language models like GPT, specifically targeting the construction of small web applications without writing code.} This tool, called NoCodeGPT, encapsulates all prompts related to technology and architectural requirements, so that the user does not need to have domain of these technical aspects. It also automatically saves the code generated by GPT in the correct folders. As a result, the user does not need to include existing code in prompts or copy GPT-generated code into local folders. Finally, it logs all interactions, prompts, and GPT responses, allowing users to easily revert to a stable version of the system in cases where the language models start producing incorrect results. In summary, with NoCodeGPT, users only write prompts related to functional requirements, while other concerns and interests are handled transparently by the proposed tool.

We also conducted two controlled experiments with the NoCodeGPT tool, in which students with limited experience in Web development were invited to use the system to develop two small Web applications (a small task management system and a question and answer forum). Despite the participants' limited experience, the results were very different from those we obtained in the first study with the standard ChatGPT interface. In the case of the first app (TodoApp), two participants implemented all four proposed user stories, while the other two participants missed just one story. In the case of the second app (ForumApp), seven participants implemented all five proposed user stories, three participants missed two stories, and one participant missed three stories. These results---entirely different from those of the first exploratory study using ChatGPT---give us confidence to assert that NoCodeGPT offers effective assistance for the construction of small Web systems by inexperienced users, without requiring the writing of any code.

The remainder of this paper is organized as follows. In Section \ref{sec::exploratory_study}, we describe our exploratory study  that motivated the construction our tool. In Section 3 we describe features and architecture of our tool. Section \ref{sec::evaluation} presents the results obtained by each experimental and the lessons learned and too threats to validity, Section \ref{sec::related_work} presents related work and finally Section \ref{sec::conclusion} concludes.

\section{Exploratory Study}\label{sec::exploratory_study}

In this section, we describe the exploratory study we initially conducted to understand the potential of ChatGPT to generate Web systems without coding, that is, solely through prompts. First, we present the methodology of this study (Section~\ref{sec::methodology}), followed by its results (Section~\ref{sec::results}). Finally, we present the main lesson learned from the study (Section~\ref{sec::lessons_learned}).

\subsection{Methodology}
\label{sec::methodology}

\noindent{\bf Reference Implementation:}
To have a reference implementation for assessing the use of ChatGPT (version 4) to generate Web apps, the first author of this paper---who is an experienced software developer--- implemented from scratch a simple Q\&A forum, which we called appForum. The rationale was to avoid evaluating an application that was possibly used by OpenAI in the training phases of ChatGPT.  We use this app as a ground-truth implementation to explore the use of ChatGPT as a nocode platform. This application implements six simple user stories:

\begin{description}
    \setlength{\parskip}{0pt}
    \setlength{\itemsep}{0pt plus 1pt}
    \item[US1:\namedlabel{US1}{US1}] As a user, I would like to register on the forum.
    \item[US2:\namedlabel{US2}{US2}] As a user, I would like to login on the forum.
    \item[US3:\namedlabel{US3}{US3}] As a user, I would like to create a question.
    \item[US4:\namedlabel{US4}{US4}] As a user, I would like to delete a question.
    \item[US5:\namedlabel{US5}{US5}] As a user, I would like to answer a question.
    \item[US6:\namedlabel{US6}{US6}] As a user, I would like to delete an answer.
\end{description}

In the back-end, we defined the implementation should use Typescript (programming language), Node.js with ExpressJS (server runtime), and SQLite (relational database). In the front-end, Vue.js and ViteJS (web frameworks) and Typescript should be used. These technologies are widely popular and they offer several advantages that make them an attractive choice for building modern web applications. In total, the back-end implemented by the first author has $565$ lines of code, three classes, and $5$ files. The relational database has three tables (\tcode{tb\_user}, \tcode{tb\_answer}, and \tcode{tb\_question}). The front-end has $820$ lines of code, $11$ files, six Vue.js components, and four pages.  Figure \ref{fig:figure-1} shows a screenshot of the main page this reference system:

\begin{figure}[H]
    \centering
    \fbox{\includegraphics[width=0.9\linewidth]{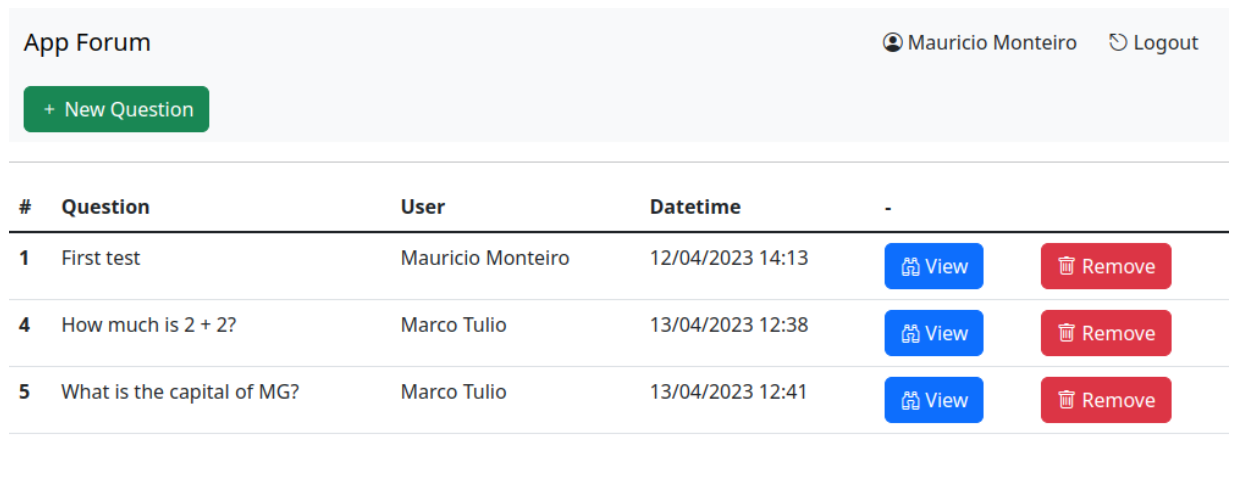}}
    \caption{Screenshot of the main screen (reference implementation)}
    \label{fig:figure-1}
\end{figure}

\begin{figure}[H]
    \centering
    \includegraphics[width=0.55\linewidth]{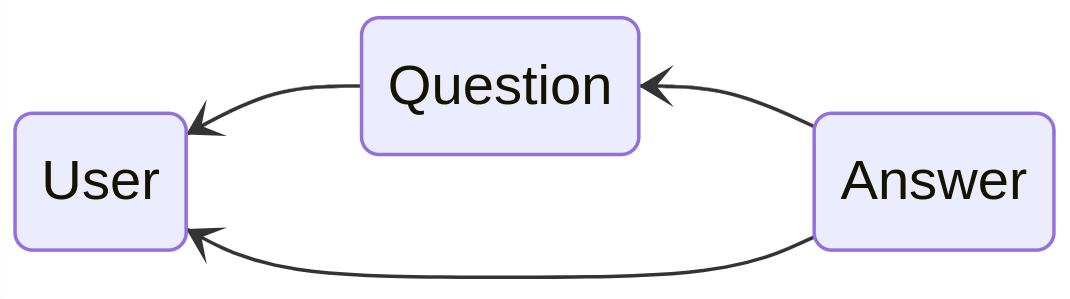}
    \caption{Class diagram (back-end, reference implementation)}
    \label{fig:figure-2}
\end{figure}

Figure \ref{fig:figure-2} shows a class diagram with the three classes of the system (\tcode{User}, \tcode{Question}, and \tcode{Answer}) and the relationship between them.

\noindent{\bf Participants:}
In the study, we asked three developers to reimplement our reference system using ChatGPT. Participant P1 has $23$ years of experience in software development. Besides being an experienced developer, he was also responsible for the reference implementation described in the previous subsection. Therefore, P1 represents the “best developer” for evaluating ChatGPT, i.e.,~we are asking an experienced developer to use ChatGPT to generate code for an application he has implemented before.\footnote{Therefore, we claim that this fact---P1 having implemented the application twice, first manually and then with the support of ChatGPT---is  something we deliberately chose to evaluate in the study.} Participants P2 is a master student in Computer Science. He also has two years of experience in software development. Finally, Participant P3 is an 4th year undergraduate CS student without previous professional software development experience.

Therefore, we attempted to recruit a diverse set of participants in terms of their software development experience and knowledge of the application to be developed with the support of ChatGPT. It is also worthnoting that the three participants had limited experience with ChatGPT, which is expected since it is a novel technology. Particularly, they have never used ChatGPT to produce a complete software application.\\[-0.3cm]

\noindent{\bf Inception Meeting:}
In this meeting, the first author presented the features, user stories, and screenshots of our appForum to both P2 and P3. Then, he asked them to rely on ChatGPT to produce code that results in an application that is as close as possible to our reference implementation and that uses the same technologies. It is important to mention that P2 and P3 had no access to the code of the reference implementation. In other words, P2 and P3 were instructed to rely only on ChatGPT to implement an identical app.\\[-0.3cm] 

\noindent{\bf Tools used by the participants:} 
Participants used only a simple text editor to copy the code generated by the OpenAI tool. Interaction was exclusively through prompts, without changing parameters such as temperature, maximum token limit, or other ChatGPT settings. Essentially, we tried to reproduce the same experience that non-experts have with language models when using the standard ChatGPT interface.\\[-0.3cm]

\noindent{\bf Review Meeting:}
After using ChatGPT separately to implement our reference system, the three participants had a series of meetings to review and assess the results achieved using the AI tool. In these meetings, they presented the prompts used to interact with ChatGPT as well as executed and discussed the code generated by the tool. They also come up with the following classification for the  prompts used in the study:

\begin{itemize}

    \item Initial Prompts: prompts that describe key functional and non-functional requirements, as well as prompts for configuring the project and installing the necessary frameworks.

    \item Feature Prompts: prompts requesting the implementation of the  features of the project, including prompts that request behaviors that were not supported by the generated code.

    \item Bug-fixing Prompts: prompts to fix bugs or incorrect behaviors in the generated code. 
    
    \item Layout Prompts: prompts to style front-end elements such as buttons, tables, and text boxes.

    \item Other Prompts:  prompts that do not fit into the previous categories, such as requesting adjustments of configurations in the developing environment. 

\end{itemize}

\subsection{Results}
\label{sec::results}

In this section, we describe the apps constructed by each participant. First, the prompts they used are summarized in Table~\ref{tab:results_participants}. \\[-0.3cm]

\begin{table}[!ht]
    \centering
    
    \begin{tabular}{p{3cm}ccc}
        
        \toprule
        {\bf\textsc{Category}} & {\bf \textsc{P1} } & {\bf \textsc{P2} }& {\bf \textsc{P3} } \\
        \midrule
        
        \textsc{Initial} & \textsc{2} & \textsc{1} & \textsc{3} \\ [0.5ex]
        \textsc{Features} & \textsc{26} & \textsc{17} & \textsc{13} \\ [0.5ex]
        \textsc{Bug Fixing} & \textsc{28} & \textsc{24} & \textsc{24} \\ [0.5ex]
        \textsc{Layout} & \textsc{7} & \textsc{9} & \textsc{0} \\ [0.5ex]
        \textsc{Other} & \textsc{2} & \textsc{2} & \textsc{6} \\ [0.5ex]
        
        \midrule
        \textsc{\textbf{Total}} & \textsc{\textbf{65}} & \textsc{\textbf{53}} & \textsc{\textbf{46}} \\ [0.5ex]
        \bottomrule
    \end{tabular}
    \caption{Prompts used by each participant}
    \label{tab:results_participants}
\end{table}

\noindent{\bf Participant \#1}: This first participant started describing the user stories and technologies adopted in the app, using the prompt:

\prompt{I need to build a Web application in TypeScript with the following programming technologies: vue@latest, Express version 4, and SQLlite3 database. The back-end should follow a stateful architecture (e.g., user ids should be stored in sessions).
The front-end should use the Bootstrap library. The app is a simple question and answer forum, which should implement the following user stories:\\
* As a user, I would like to register on the forum.\\
* As a user, I would like to login on the forum.\\
* As a user, I would like to create a question.\\
* As a user, I would like to delete a question.\\
* As a user, I would like to answer a question.\\
* As a user, I would like to delete an answer.}{P1 participant's initial prompt.}\\

Despite listing the user stories in the initial prompt, P1 had to use $26$ other prompts to request refinements in the code generated by ChatGPT, such as in the following prompt:

\prompt{Please, create a new page such that the user can view the answers for a selected question. In the main page, add a “View Answers" button that will then present this new page.}{Another initial prompt used by participant P1}{}

As another example, P1 had to elaborate prompts explicitly requesting the record of new answers to a particular question. He also used other prompts to implement missing user stories and to include missing information, e.g., the name of the user who answered each question.

\prompt{In the Answers Preview Screen, we should be able to register an answer to a question.}{Feature prompts to register new answer}{p:first}{}

Moreover, P1 used $28$ prompts to fix bugs in the code produced by ChatGPT, such as: 

\prompt{There is an error in the browser console:
Uncaught SyntaxError: ambiguous indirect export: `setAuthenticated'.
Could you fix it?}{Bug Fixing Prompt used by P1 for to fix the code produced by ChatGPT}{}

Finally, P1 used seven layout prompts. 
For example, in the code initially produced by ChatGPT the data about questions (ID, title, etc) was presented in the page as items of a list and not as rows of a table, as planned by the participant. Thus, P1 used the following prompt to request the correct layout:

\prompt{I would like the `client/src/views/Questions.Vue' file responsible for registering questions to list the questions in a table with the question ID, the question title and the commands to view the answers to the question and delete the question.}{Layout Prompt used by participant P1}{}

\begin{figure}[!ht]
    \centering
    \fbox{\includegraphics[width=0.7\linewidth]{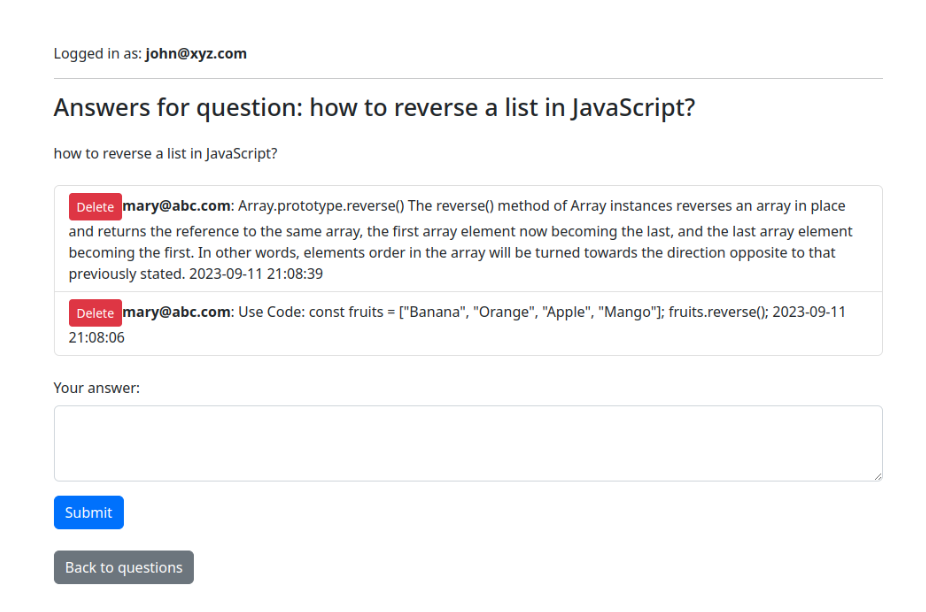}}
    \caption{Screenshot of the answers screen (Participant P1)}
    \label{fig:figure-4}
\end{figure}

Figure \ref{fig:figure-4} shows a screenshot of the app implemented by P1 using ChatGPT. The page presented in this figure is used to post answers.

\vspace{1em}

\noindent{\bf Participant \#2:}
In this case, only one initial prompt was needed, as follows: 

\prompt{I would like to create an user authentication form web page, with the fields ``e-mail" and ``password", with two green buttons: ``sign up", which will redirect the users to a new form page where he can sign up, and ``sign in", which will authenticate existing users and redirect them to a home page, returning an error when the user is not registered. I want to use the following technologies for this: Vue.js, Express.js, TypeScript and Sqlite. Could you help me with that, from installing these technologies to build those pages.
}{P2 participant's initial prompt.}{}

As the reader may notice, P2 started by requesting the implementation of an specific feature (authentication), by providing a high level text describing the main fields and buttons, complemented with a request for basic error handling when the user does not exist, and then explicitly listing the desired technologies for the project. In the end, he asked ChatGPT to help in the whole process, from installing the required technologies to build the specified page.

As a response to this initial prompt, ChatGPT recommended to divide the problem in smaller steps and provided guidelines to each one (e.g., separating code by domain and indicating where newer code from an existing domain should be added). After everything was configured, ChatGPT suggested the code for the sign up/sign in feature.

However, P2 figured out that ChatGPT's answer was not fully correct and functional, which required him to write another 17 prompts to fix bugs in the generated code. As examples of such bugs,  we can mention libraries that were actually not installed or not imported where needed, the back-end code was not correctly integrated with the front-end, and there were missing parameters on the project's configuration files. 


For the remaining features, P2 changed his strategy and asked ChatGPT to first generate the front-end code with mocked objects. This first version of the front-end was then carefully tested. After that, it was integrated with the back-end code. For this last step, P2 decided to use prompts that include both the front-end and the current back-end code. Such prompts requested ChatGPT to extend the back-end with logic to handle the new features that were previously implemented and tested in the front-end. As an example of this new strategy, we have the following prompt, where P2 requests ChatGPT to generate code in the back-end (file \tcode{app.ts}) to persist an answer available in a form in the front-end (file \tcode{PostDetails.vue}). 

\prompt{Now, I want to get the post's answers available in PostDetails.vue and, when submitting the form with the answer, I want it to be saved in the database. My code from PostDetails.vue looks like that: $[$Source code from PostDetails.vue file$]$ and my app.ts looks like that: $[$Source code from app.ts file$]$ \\How can I do that?}{Prompt example of P2 asking ChatGPT to generate code in the backend to persist an answer available in a form in the frontend.}{}

Regarding the style prompts, P2 was able to obtain the desired style by describing how the elements should look like, for example by informing the hexadecimal code of particular elements' colors or the shape of buttons (e.g., a button with a stadium-shaped border). He also experienced naming the colors and asking for darker or clearer tones of existing ones, which ChatGPT understood as well. 

\begin{figure}[H]
    \centering
    \fbox{\includegraphics[width=0.95\linewidth]{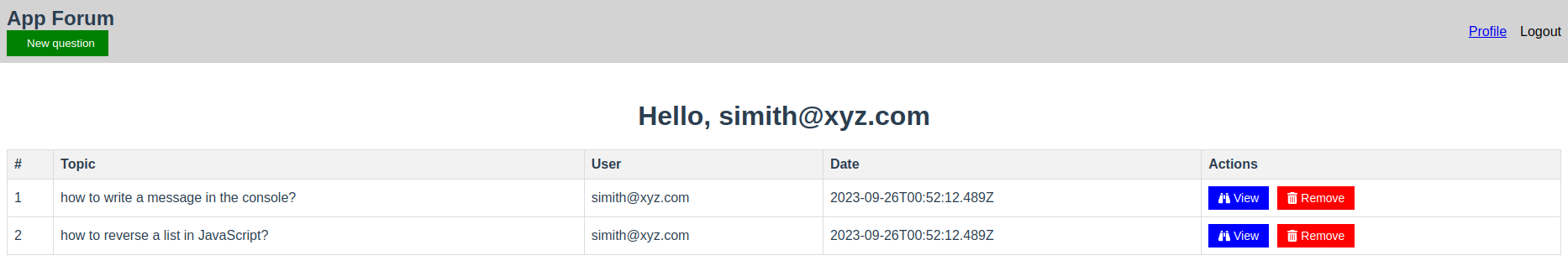}}
    \caption{Screenshot of the question list page (Participant P2)}
    \label{fig:figure-6}
\end{figure}

Figure \ref{fig:figure-6} shows a screenshot of the page to visualize questions and to provide answers, as implemented by P2 using ChatGPT.

\noindent{\bf Participant \#3:} This participant started with the following prompt that describes the technologies and basic features of the application and asks ChatGPT to build the code:

\prompt{Please, build a forum app with the following technologies: Vue.js, Express.js, TypeScript, and Sqlite. The app should have a login screen and an option to register a user if he/she haven't already. After logging in, the user should have the option to create a Question and to see the questions that have been created before. By clicking on the ``Answer” button, a new screen will appear and the user will be able to provide an answer, save it, and then return to the main screen.}{P3 participant's initial prompt.}{}

The answer of this first prompt included a good portion of the project files and also instructions to install the required technologies. However, two more prompts were necessary to obtain details about such instructions, including the following prompt:

\prompt{In topic 5 (``Set up the front-end"), how do I create views for each route?}{Example of prompts to obtain details about such instructions.}{}

After these initial prompts, P3 asked ChatGPT to implement other features by using 13 prompts, such as:

\prompt{Please implement code to make API calls to the back-end server for user registration, fetching questions, and saving new questions.}{Feature prompts used by participant P2.}{}

However, the code generated by ChatGPT missed some important files, such as index.js (in
the backend). Indeed, ChatGPT informed that it was necessary to implement this file, but it does
not provided the code, even after an explicit request.
As a result, several bugs persisted in the generated code. Moreover, important user stories such as user registration and login were not properly implemented. The front-end for such features was created but it was not able to call the correspondent code in the back-end. ChatGPT correctly attributed this problem to an error in the connection between the front-end and back-end. However, when P3 attempted to fix the bug, ChatGPT entered in a loop, continuously suggesting previous (and also incorrect) versions of the code.  After 18 attempts, P3 concluded that it was not possible to advance and he decided to quit with the project not finished.

\subsection{Lesson Learned}
\label{sec::lessons_learned}

The main lesson learned from this study is the importance of the developer's proficiency in the technologies and frameworks used by the target system. For instance, the first two participants—who successfully completed the proposed application—were well-versed in web development and its associated technologies and frameworks. As a result, they were able to leverage their experience to formulate prompts that guided ChatGPT in fixing the bugs in the code generated by the tool.\\[-0.3cm]

\vspace{1em}
\lessonlearned{
{\bf Lesson Learned}: When using the standard interface provided by ChatGPT to support the end-to-end construction of web apps, experience in software development practices, architectures, and technologies is crucial. This expertise is especially important when formulating bug-fixing prompts. In other words, it is unrealistic to expect non-developers to write these prompts and create a functional web app using ChatGPT.}

\subsection{Threats to Validity}
\label{sec::threats_to_validity-exploraty-study}

There are two main threats to the validity of the results reported in this exploratory study. First, our reference application (a question and answer forum) may not represent the universe of systems that are built from scratch by software developers. However, we chose a well-known application that follows a common architecture  and that uses popular technologies.
Second, the code of the reference application was generated by ChatGPT with prompts formulated by three developers. Therefore, these participants may not represent the universe of developers who intend to use ChatGPT to support end-to-end software construction. However, we selected developers with diverse profiles and levels of experience in software development.

\section{NoCodeGPT: Features and Architecture}
\label{sec::nocodegpt}

Based on the lessons learned from our exploratory study, we decided to implement a new interface for GPT, with specific features for building small web apps using only prompts (i.e., without writing any code). This interface, called NoCodeGPT, can act as a replacement for the traditional interface (ChatGPT). The idea is to enable not only developers (like participants P1 and P2 from the exploratory study) to use language models for generating web applications. Instead, we intend that users with limited software development experience (like participant P3) could also be able to generate small web applications without having to write a single line of code.

In this section, we first describe the key features of NoCodeGPT (Section~\ref{sec:nocodepgt-featues}). These features emerged from our experience and observation in the Exploratory Study. Next, we also present the architecture and implementation of NoCodeGPT (Section~\ref{sec:architecture_implementation}).

\subsection{Main Features}
\label{sec:nocodepgt-featues}

The main functionalities of the tool are as follows:\\

\begin{figure}[H]
    \centering
    \fbox{\includegraphics[width=0.95\linewidth]{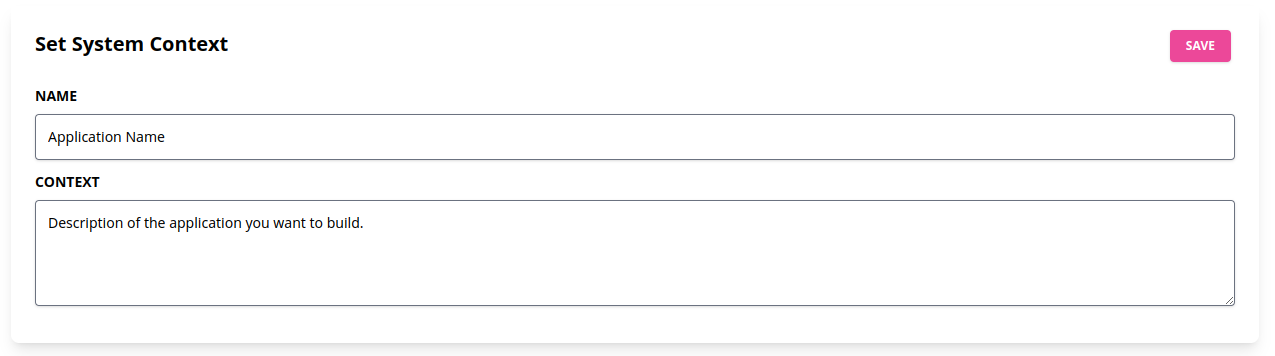}}
    \caption{Page to enter the name and main context of the system under development}
    \label{fig:figure-set-context}
\end{figure}

\noindent {\bf Initial Prompts:}
The tool defines internally the initial prompts for building web systems, meaning the prompts that specify the system's technology, architecture, main directories, among other decisions. This way, the user does not need to create or provide these prompts. 
Additionally, NoCodeGPT requests that the user provide a prompt describing the core functionality of the system under construction (see Figure \ref{fig:figure-set-context}). This prompt is important for establishing the contextual framework to be considered by the GPT model when generating the system.

\noindent{\bf Pre-defined features:} \label{sec::pre-defined-features}
Some features are very common in web system. Therefore, we decided to provide built-in and pre-tested prompts to automatically generate and include such features in the systems generated by NoCodeGPT. Currently, the tool supports two pre-defined features: a login and a user registration page.\\

\noindent{\bf Creation and Refinement of Pages:}
Since NoCodeGPT is exclusively designed to build web systems, it assumes that these systems consist of a set of pages. For each page, the user must provide a name and a brief description of the desired functionalities. For example, in the case of a Q\&A forum, the user might create a page called \textit{Question Submission} with the following description: ``I would like to be able to submit questions on this page. For each question, I would like the system to store the following information:"

\begin{figure}[H]
    \centering
    \fbox{\includegraphics[width=0.95\linewidth]{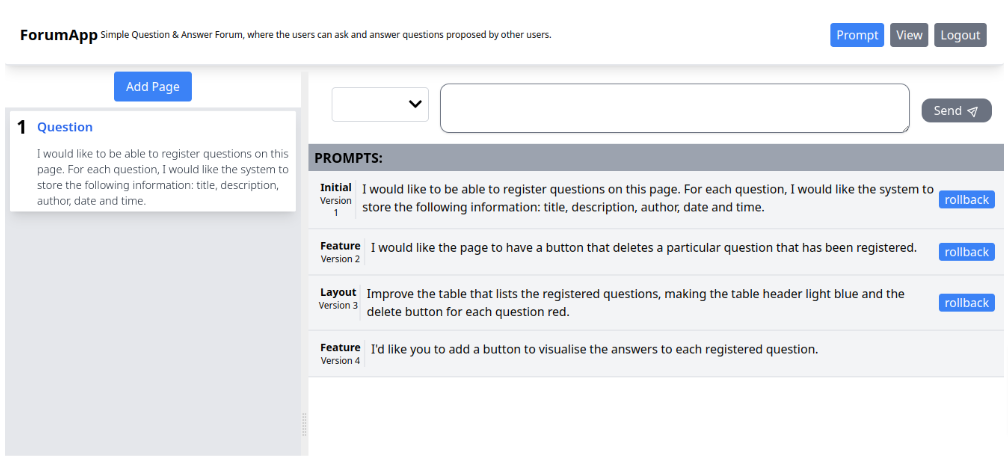}}
    \caption{Prompts for refining generated pages}
    \label{fig:figure-NoCodeGPT-tool}
\end{figure}

However, as we clearly learned in the Exploratory Study,
it is unlikely that the GPT model will generate the ideal page the user envisions in the first iteration. Consequently, NoCodeGPT includes a refinement feature where the user can request improvements to a specific page.  These prompts can be of three types: new features (such as, \textit{the page should have a button to delete a question}); bug fixes (such as, \textit{the delete question button is not working}), and layout adjustments (such as, \textit{the delete question button should be positioned right after the question's title}).
Figure \ref{fig:figure-NoCodeGPT-tool} shows an example of this functionality for the Q\&A Forum. In this figure, the user is refining the \textit{Question Submission} page.

Since refinements are restricted to one page, the tool's implementations is able to add context to the prompts provided by users. For example, our implementation automatically adds the source code files that need to be refined by the GPT model in such prompts. On the one hand, this functionality is essential to allow the tool to be used by inexperienced users, as they usually cannot correctly identify the source code files in which they intend to make a change. On the other hand, it increases the effectiveness of language models by providing them with the precise code they need to work on.\\

\noindent{\bf Web Page Transitions:}
Another interesting aspect is a feature provided by NoCodeGPT to connect two pages. For instance, in our forum, the user must independently create the \textit{Question Submission} and \textit{Answer Submission} pages. However, afterward, they must return to the \textit{Question Submission} page and request the creation of a button on each question that shows the page with its answers. An example is shown below:

\prompt{Add a button to view the registered answers on the Answers page.}{Example that the tool uses to "connect" two pages.}{}

After various tests, we concluded that this alternative using simple prompts is the best solution for connecting pages. That is, the user should first create and test the pages individually. Then, they should  request the connections between the pages, whenever necessary. To do this, they should simply use a prompt asking that when a certain component is clicked another page should be displayed.\\

\noindent{\bf Version Management:}
NoCodeGPT also implements a simplified version control system for the code generated with the assistance of GPT. This allows users to easily restore a previous version if GPT starts generating invalid code that does not comply with the system specifications. For example, suppose the user formulates prompts requesting the implementation of features $F_1, F_2, \dots, F_n$, with each submission generating a respective version $V_i$ of the system. However, in a particular version $V_j$, the language model might start hallucinating and generate invalid code. In such cases, the user can revert to version $V_{j-1}$ using a single button provided in NoCodeGPT's interface.

This feature was specifically designed to address a difficulty encountered by Participant P3 in our Exploratory Study. This participant abandoned the study because ChatGPT chose a path that resulted in successive invalid versions of the system. Moreover, using the standard ChatGPT interface, backtracking to the last correct version was not easy. He would have had to save these versions manually, which is not a natural procedure for less experienced users.\\

\noindent{\bf Execution and Visualization:} NoCodeGPT also provides a button that allows the user to quickly run and check the behavior of the code generated by GPT. This way, the user does not need to have knowledge of command-line tools, such as compilers and interpreters.

\subsection{Architecture and Implementation} \label{sec:architecture_implementation}

\begin{figure}[ht]
    \centering
    \fbox{\includegraphics[width=0.75\linewidth]{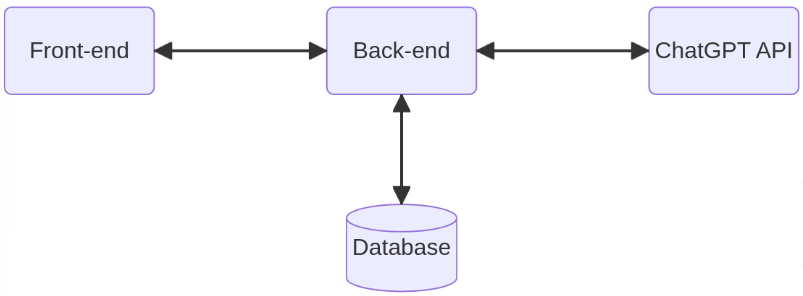}}
    \caption{NoCodeGPT's high-level architecture}
    \label{fig:figure-NoCodeGPT-diagram}
\end{figure}

The implementation of NoCodeGPT has two modules: front-end and back-end, as shown in Figure \ref{fig:figure-NoCodeGPT-diagram}. The front-end is responsible for controlling the functionality of the tool's pages and for guiding the user through the workflow to build the Web app. The back-end is responsible for receiving requests sent by the front-end and processing the necessary information to build the prompts that are submitted to the OpenAI API. The back-end also stores all interactions with the OpenAI platform in a database.

Specifically, the front-end implementation uses the following technologies: Vue.js (version 3) with TypeScript using the Composition API, HTML, and CSS. For the back-end implementation, ExpressJS, SQLite 3, and JSON Web Token are used. The front-end has 1,847 lines of code, and the back-end has 3,342 lines, totaling 5,189 lines.

\begin{figure}[!ht]
    \centering
    \fbox{\includegraphics[width=0.75\linewidth]{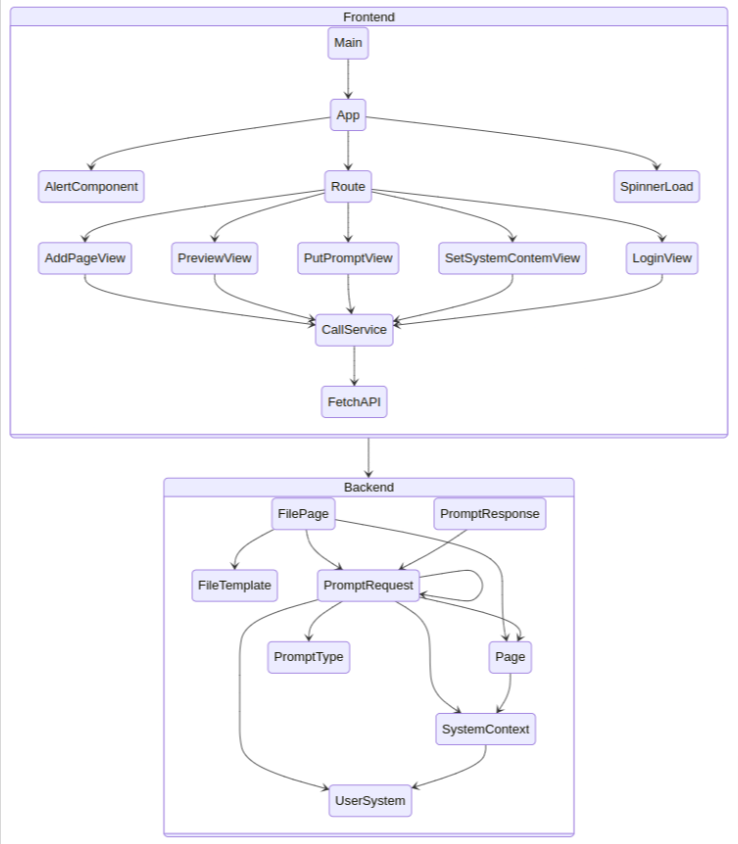}}
    \caption{Front-end and Back-end architecture}
    \label{fig:NoCodeGPT-classes}
\end{figure}

Figure \ref{fig:NoCodeGPT-classes} provides a detailed view of the architecture of the implemented tool. The front-end consists of the \tcode{Main} class, which controls the initialization of this component, and the \tcode{App} class, which manages the interface logic. Page route management is handled by the \tcode{Route} class, which is capable of displaying any of the pages, such as \tcode{AddPageView} and \tcode{LoginView}. Specifically, the application pages are implementations of Vue Single-File Components. These components call the endpoints implemented in the back-end using the \tcode{CallService} class, which relies on the standardized implementation of the FetchAPI class.

In the back-end, the prompts are initially processed by adding other important information to obtain more accurate responses and, ultimately, the GPT API is called to generate the code requested by users. The \tcode{PromptRequest} and \tcode{PromptResponse} classes store, respectively, the requests and responses made to this API. Detailed information from the previous requests is also stored to implement the version control mechanism provided by the tool.
Finally, the \tcode{PromptType} class stores the templates of the prompts sent to the OpenAI platform.\\ 

\noindent{\bf Temperature:} From the experience of the Exploratory Study, we concluded that lowering the temperature in fact makes the model more deterministic and also accurate. Therefore, we decided to reduce the temperature used in NoCodeGPT to zero.

\section{NoCodeGPT Evaluation}
\label{sec::evaluation}

In this section, we present an evaluation of NoCodeGPT, conducted in two phases, with a total of 14 participants, who were invited to implement two small web applications using only prompts and our tool. Our main objective is to verify whether NoCodeGPT helps to overcome the limitations identified in the Exploratory Study.

\subsection{Methodology}

The evaluation of the proposed tool was conducted in two phases. First, a pilot test was carried out with four participants. The goal was to test and validate the implementation of the proposed tool with a small number of participants and assess its feasibility. Specifically, the participants used the tool to implement a simple task management application with the following user stories:

\begin{enumerate}[itemsep=0pt]
\item As a user, I want to add a new task to the list so I can track what I need to do.

\item As a user, I want to edit an existing task so I can correct or update the task description.

\item As a user, I want to mark a task as completed so I can indicate that I have finished it.

\item As a user, I want to remove a task from the list so I can clean up my todo list.
\end{enumerate}

The participants received initial training on the tool (30 minutes) and were also provided with a low-fidelity mockup of the application. We decided to provide this mockup so that participants could have an idea of the functionality they needed to implement. However, we clarified that the mockup would serve only as an example, meaning the application to be created did not need to have an identical layout.

Following this pilot experiment, we conducted a larger experiment with 10 participants. In this case, we used the ForumApp from our Exploratory Study (Section~\ref{sec::exploratory_study}) as the target application. The participants also received initial training and a mockup of the requested app. They were asked to implement the following user stories:

\begin{enumerate}[itemsep=0pt]
\item As a user, I want to add a new question to the forum.
    
\item As a user, I want to delete a question from the forum.

\item As a user, I want to access the answers of a given question in a separated page.

\item As a user, I want to answer to a question in the forum.

\item As a user, I want to delete an answer.
\end{enumerate}

Compared to the Exploratory Study, we made two important changes to the list of user stories of ForumApp. First, we removed the stories related to login and user registration, as these are predefined features in NoCodeGPT, as explained in Section~\ref{sec::pre-defined-features}. Second, we added a story explicitly stating that, given a question, it is important to access a second page with its answers. This was done to prevent users from implementing the entire app on a single page. If that happened, we would not  be able to evaluate NoCodeGPT’s ability to support the creation of apps composed of multiple web pages.

Table \ref{tab:year_of_experience} summarizes the participants' web development experience in each experiment (TodoApp and ForumApp). All participants are undergraduate Computer Engineering students in their first or second year. Additionally, as shown, they match the ideal user profile for NoCodeGPT, meaning they have either no experience or, at most, one year of experience in web development.

\begin{table}[!ht]
    \centering
    \begin{tabular}{p{7cm}cc}
        
        \toprule
         {\bf\textsc{Years of experience}} & {\bf \textsc{TodoApp}} & {\bf \textsc{ForumApp}} \\
        \midrule
        
        No experience & \textsc{2} & \textsc{8} \\ [0.5ex]
        Less than one year's experience & \textsc{2} & \textsc{2} \\ [0.5ex]
        
        \bottomrule
        
    \end{tabular}
    \caption{Participants’ experience on web development}
    \label{tab:year_of_experience}
\end{table}

\subsection{Pilot Experiment Results (TodoApp)}
\label{sec::evaluation_todoapp}

\begin{figure}[!ht]
    \centering
    \fbox{\includegraphics[width=0.65\linewidth]{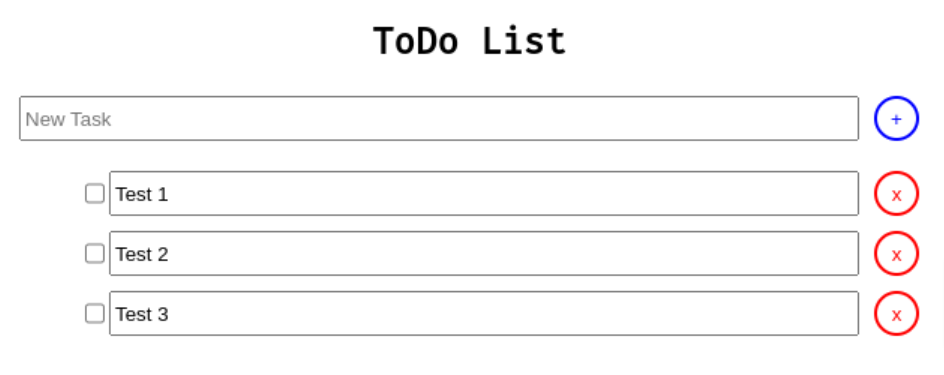}}
    \caption{Screenshot of participant P2's application.}
    \label{fig:figure-NoCodeGPT-example-app}
\end{figure}

In the pilot experiment, two participants managed to build the TodoApp application with all four proposed stories. However, two participants did not implement one of the stories. Figure \ref{fig:figure-NoCodeGPT-example-app} shows a screenshot of an application built by one of the participants who successfully completed all four stories.

Moreover, Table \ref{tab:user_stories_todo_app} details the stories that were implemented by the pilot participants. As we can see, participant P4 was unable to implement task editing, while participant P1 was unable to implement a feature to mark a task as completed. In both cases, the participants attempted to implement the functionality, but the code generated by GPT did not work as expected. Thus, they decided to give up without attempting again. On the other hand, the creation of new tasks, as well as their removal, were implemented by all four participants.

\begin{table}[!ht]
    \centering
    \begin{tabular}{p{8cm}c}
        
        \toprule
         {\bf\textsc{User Stories}} & {\bf \textsc{Participants who succeeded}} \\
        \midrule
        
        As a user, I want to add a new task & \textsc{P1, P2, P3, P4}  \\ [0.5ex]
        As a user, I want to edit an existing task & \textsc{P1, P2, P3} \\ [0.5ex]
        As a user, I want to mark a task as completed & \textsc{P2, P3, P4} \\ [0.5ex]
        As a user, I want to remove a task from the list & \textsc{P1, P2, P3, P4} \\ [0.5ex]
        
        \bottomrule
        
    \end{tabular}
    
    \caption{Participants who implemented each user story in the pilot experiment (TodoApp).}
    \label{tab:user_stories_todo_app}
\end{table}

\begin{figure}[H]
    \centering
    \fbox{\includegraphics[width=0.65\linewidth]{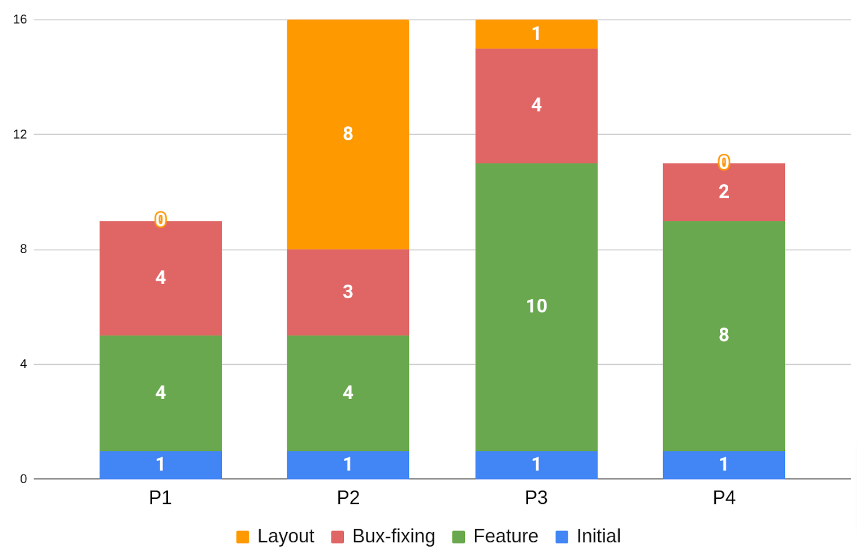}}
    \caption{Types of prompts implemented by each participant in the pilot experiment (TodoApp).}
    \label{fig:figure-NoCodeGPT-graph1}
\end{figure}

Figure \ref{fig:figure-NoCodeGPT-graph1} shows the number of prompts each participant used to build the TodoApp. Participants P2 and P3 interacted the most with the tool and were the only ones to implement all user stories, each using 16 prompts. In contrast, participants P1 and P4 used nine and eleven prompts, respectively. The most common prompt type was for requesting features, except for participant P2, who used eight prompts to improve the app's layout. For P1, there was an even split between feature and bug-fix prompts, with four prompts each. Interestingly, P1 and P4 did not use any layout prompts, indicating they were satisfied with the initial layout proposed by GPT. Lastly, all participants used exactly one initial prompt.

NoCodeGPT's version control functionality was heavily used during the pilot experiment. The four participants carried out a total of 16 rollbacks while building the proposed web application. Table \ref{tab:total_rollbacks1} indicates that participant P4 made the most use of this feature, requesting five rollbacks.

\begin{table}[!ht]
    \centering
    \begin{tabular}{lcccc}
        
        \toprule
        
        {\bf \textsc{Participant}} & {\bf \textsc{P1}} & {\bf \textsc{P2}} & {\bf \textsc{P3}} & {\bf \textsc{P4}}   \\
        \midrule
        
        Rollbacks & \textsc{4}  & \textsc{4} & \textsc{3} & \textsc{5} \\ [0.5ex]
        
        \bottomrule
        
    \end{tabular}
    
    \caption{Rollbacks by participants in the pilot experiment (TodoApp).}
    \label{tab:total_rollbacks1}
\end{table}

\begin{figure}[!ht]
    \centering
    \fbox{\includegraphics[width=0.85\linewidth]{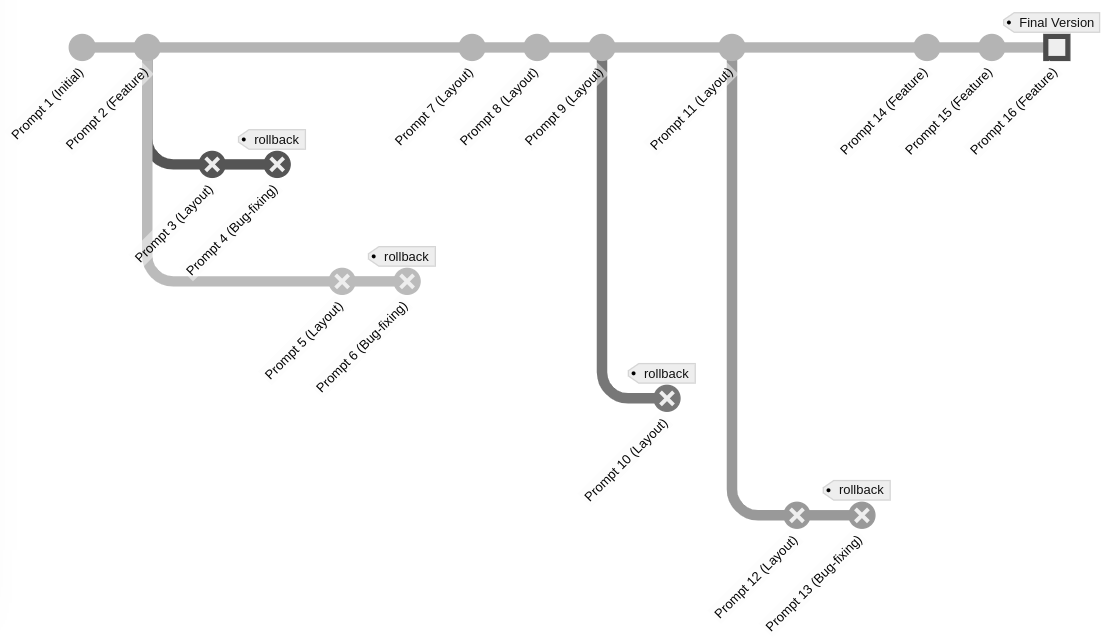}}
    \caption{Rollbacks performed by Participant P2 in the pilot experiment (TodoApp).}
    \label{fig:figure-NoCodeGPT-rollback1}
\end{figure}

Figure \ref{fig:figure-NoCodeGPT-rollback1} illustrates the rollback operations carried out by participant P2. In this figure, each branch represents a sequence of prompts that were abandoned (i.e., that were concluded with a rollback). Specifically, P2 performed four rollbacks, after prompts 4, 6, 10, and 13. We can also see that P2 used 16 prompts in total, but seven prompts were discarded. Only nine prompts resulted in usable code (prompts 1-2, 7-9, 11, 14-16).

\subsection{Results of the Second Experiment (ForumApp)}
\label{sec::evaluation_forumapp}

\begin{figure}[!ht]
    \centering
    \fbox{\includegraphics[width=0.85\linewidth]{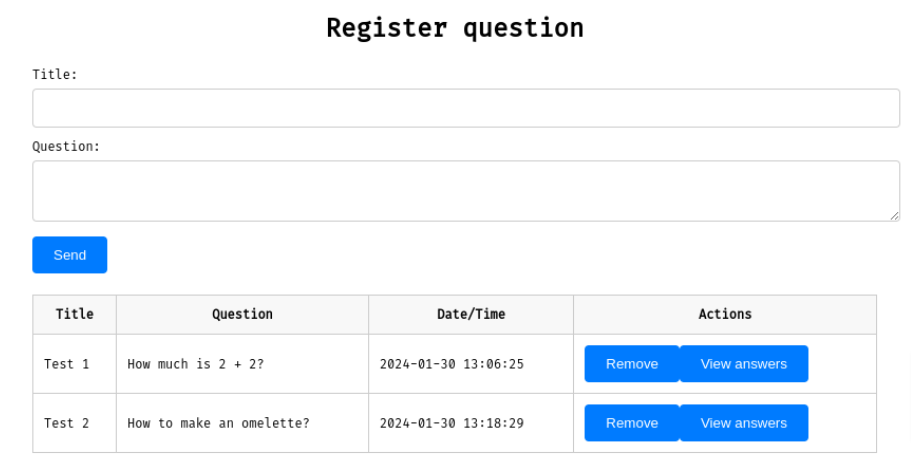}}
    \caption{Question registration page created by participant P7 in the second experiment.}
    \label{fig:figure-NoCodeGPT-example2}
\end{figure}

\begin{figure}[!ht]
    \centering
    \fbox{\includegraphics[width=0.85\linewidth]{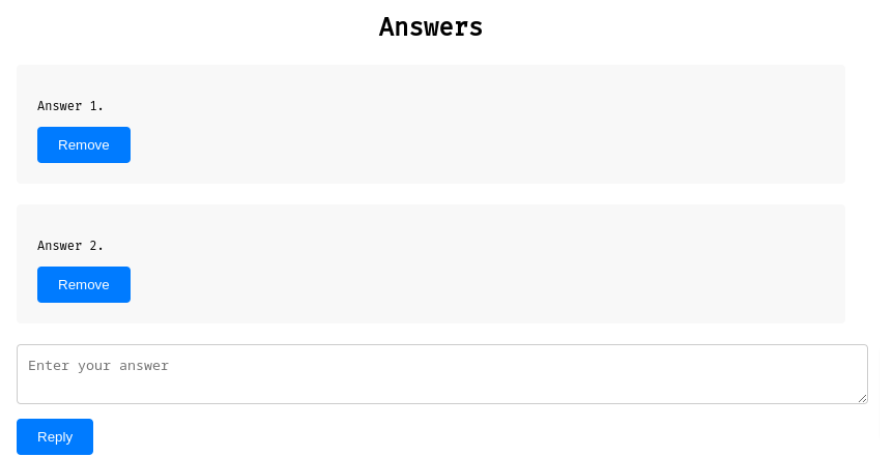}}
    \caption{Answer registration page created by participant P7 in the second experiment.}
    \label{fig:figure-NoCodeGPT-example3}
\end{figure}

In the second experiment (ForumApp), seven participants successfully implemented all five stories. The last three participants failed to implement at least two user stories. Figures \ref{fig:figure-NoCodeGPT-example2} and \ref{fig:figure-NoCodeGPT-example3} show screenshots of the application created by one of the participants who completed all five stories. Figure \ref{fig:figure-NoCodeGPT-example2} shows the question registration page, and Figure \ref{fig:figure-NoCodeGPT-example3} shows the answer registration page.

Table \ref{tab:user_stories_forum_app} shows the stories implemented by the participants in the second experiment. As we can see, participants P8, P9, and P10 were unable to implement the stories requesting answering questions and deleting answers. Participant P9 was the only one who failed to implement the third story, which defines that from the question page it should be possible to access the answers pages.

\begin{table}[!ht]
    \centering
    
    \begin{tabular}{p{8cm}c}
        
        \toprule
        {\bf\textsc{User Stories}} & {\bf \textsc{Participants who succeeded}}  \\
        \midrule
        
        As a user, I want to add a new question &  All participants  \\ [0.8ex]
        As a user, I want to remove an existing question & All participants \\ [0.8ex]
        As a user, I want to access the answers page & All participants, except P9 \\ [0.8ex]
        As a user, I want to answer a question & All participants, except P8, P9, P10 \\ [0.8ex]
        As a user, I want to delete an answer & All participants, except P8, P9, P10 \\ [0.8ex]
        
        \bottomrule
        
    \end{tabular}

    \caption{Participants who implemented each user story in the second experiment}
    \label{tab:user_stories_forum_app}
\end{table}

\begin{figure}[ht!]
    \centering
    \fbox{\includegraphics[width=0.75\linewidth]{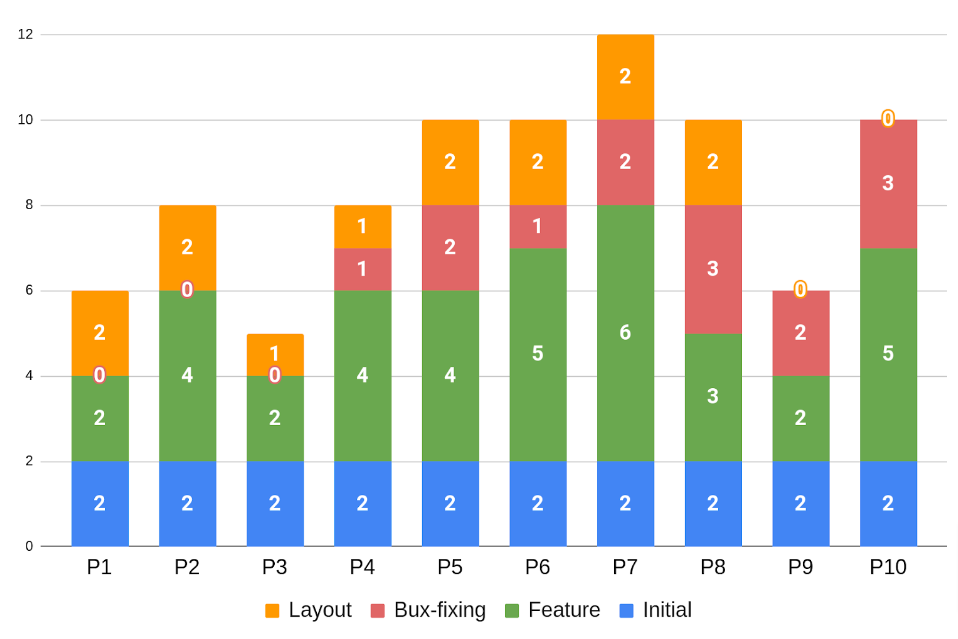}}
    \caption{Total prompts used by each participant stacked by type.}
    \label{fig:figure-NoCodeGPT-graph2}
\end{figure}

Figure \ref{fig:figure-NoCodeGPT-graph2} shows the number of prompts used by participants to build the ForumApp. Feature-type prompts were the most frequently used by participants, totaling 37 prompts, followed by 20 initial-type prompts. Each participant used two initial-type prompts: one for the page for registering questions and another for the page for registering answers. The bug-fixing and layout categories have 14 prompts each.

Of the seven participants who successfully implemented the application, P7 used the most prompts (12), while P3 used the fewest (5).
The main reason for this difference seems to be the level of detail in their initial prompts. P3 started with a more detailed prompt, as presented next:

\prompt{Create a page where I can register questions, where the registration includes a title, the text of the question, the logged-in user who asked the question, and the time the user asked. The questions should be listed in a table showing the title, the question text, the user who asked, and the date/time, alongside a button to delete the question and a button to view answers. The view answers button should go to an answer registration page for the selected question.}{}{}

On the other hand P7's initial prompt was more succint, as follows:
 
\prompt{Create a question registration page, which will have a box to enter the title of the question and add the description of the question. The questions will be listed separately in a table that will show the title of the question, the user who registered it, and the date and time.}{}{}

Participants P9 and P10 were the only ones who did not use layout-type prompts. This behavior is common among participants who struggle to finalize the application. Since both participants were unable to implement all functionalities, they did not reach the stage where layout-type prompts are typically used, i.e., after everything is working as expected. Participants P1, P2, and P3 were the only ones who did not use bug-fixing prompts, likely because their more detailed initial prompts led to more precise results by the GPT model.

Finally, table \ref{tab:total_rollbacks2} shows the number of rollbacks by each participant. As we can see, this feature was used less during the development of ForumApp (compared to the pilot study with TodoApp). Our hypothesis is that even though ForumApp has two pages, they are simpler than the single page specified in TodoApp. These simpler pages make it easier for the participants to define more precise and effective prompts. As a result, GPT was able to generate correct code in the first attempts and the need for rollbacks decreased.

\begin{table}[!ht]
    \centering
    
    \begin{tabular}{lcccccccccc}
        
        \toprule
        {\bf\textsc{Participant}} & {\bf \textsc{P1}} & {\bf \textsc{P2}}  & {\bf \textsc{P3}}  & {\bf \textsc{P4}}  & {\bf \textsc{P5}}  & {\bf \textsc{P6}}  & {\bf \textsc{P7}}  & {\bf \textsc{P8}}  & {\bf \textsc{P9}}  & {\bf \textsc{P10}}   \\
        \midrule
        
        Rollbacks & \textsc{0} & \textsc{0} & \textsc{0} & \textsc{0} &  \textsc{0} &  \textsc{1} & \textsc{2} & \textsc{0} & \textsc{0} & \textsc{2} \\ [0.5ex]
        
        \bottomrule
        
    \end{tabular}
    
    \caption{Rollbacks by participants in the second experiment.}
    \label{tab:total_rollbacks2}
\end{table}

\subsection{Participants' Perceptions}

After the experiment, the participants of the first and second experiment (in total, 14 participants) also reported their perceptions about NoCodeGPT in a simple form. They answered two questions:\\[-0.3cm]

\noindent{\bf What are the tool's most positive points?}
Among the positive points highlighted by the participants were the simplicity of NoCodeGPT's interface (10 participants), the convenience of not having to copy-and-paste code (9 participants), and the lack of need for prior knowledge of programming languages (10 participants). Additionally, the ability to revert to previous versions and quickly view the prototype was seen as a valuable feature for those developing an application. Below are some of the positive aspects of the tool mentioned by the participants:\\[-0.25cm]

\noindent \textit{We were able to create a functional application without using any code, just simple instructions. The possibility of selecting previous versions was fundamental for correcting some bugs.}\\[-0.25cm]

\noindent \textit{NoCodeGPT has an interesting purpose and is very useful, especially if you have a minimum knowledge of development. One of the best features is being able to go back to previous versions.}\\[-0.25cm]

\noindent \textit{The tool is very easy to use; and the feature of not having to copy-and-paste any code or have prior knowledge of programming languages (although knowledge certainly increases is benefit); The feature to go back a version is very useful and helps a lot in the construction.} \\[-0.25cm]

\noindent{\bf What are the downsides of the tool? }
The main negative point pointed out by participants was the delay in responses during interactions with the prompts (14 participants), as commented by the following participants:\\[-0.25cm]

\noindent \textit{NoCodeGPT at times takes too long to respond. Improve the response time.} \\[-0.25cm]

\noindent  \textit{Slow responses, I suggest increasing the processing speed of the tool.} \\[-0.25cm]

\noindent \textit{It takes too long to execute the commands.} \\[-0.25cm]

However, it is important to note that this delay refers to accessing the GPT API provided by OpenAI. Thus, it is a variable over which we have no control.

\subsection{Lessons Learned}

In the exploratory study, we concluded that the default interface offered by ChatGPT is not suitable for users without programming experience. Therefore, we decided to invest in the design and implementation of NoCodeGPT, a wrapper for the GPT API with features that make it easier for non-expert users to create a simple web application without writing code.

In this section, we reported the results of two studies designed to evaluate the effectiveness of NoCodeGPT in accomplishing its purpose. In the first study (TodoApp), two participants implemented all four proposed user stories, while the other two participants missed just one story. In the second study (ForumApp), seven participants implemented all five proposed user stories, three participants missed two stories, and one participant missed three stories.\\

\lessonlearned{
{\bf Lesson Learned:}
The results obtained with both apps showed that NoCodeGPT is effective in helping inexperienced users implement simple web applications without having to write code.

\begin{itemize}[itemsep=0pt]

\item In the exploratory study, using the default interface provided by ChatGPT, an inexperienced participant was unable to implement any of the proposed user stories.

\item In the experiments described in this section, using the customized interface provided by NoCodeGPT, a scenario of total failure did not occur with any of the participants. Considering both experiments, NoCodeGPT was used by 14 inexperienced participants. More than half of such participants (9 participants) successfully completed the proposed applications, while the others completed at least half of the proposed user stories.
\end{itemize}
}

\vspace{0.1cm}
Next, we briefly discuss the role and contribution that the main features of NoCodeGPT had in these results.

\begin{itemize}[itemsep=0pt]

\item NoCodeGPT establishes a clear distinction between two categories of prompts: (1) prompts that define technologies and architectures; and (2) prompts that define functional requirements. The latter prompts are handled and encapsulated by the proposed tool. Consequently, users are solely responsible for writing prompts describing functional requirements.

\item NoCodeGPT also manages and stores the code generated by GPT. This way, users do not need to copy and paste the generated code to a local directory or create a specific folder structure for each application and architecture. These concerns are completely automated by the proposed tool. It also automates related tasks, such as installing libraries and configuring environment variables. It is worth mentioning that such tasks represented major obstacles for the non-expert participant from our Exploratory Study (Section~\ref{sec::exploratory_study}).

\item The rollback feature has been of major importance in NoCodeGPT's performance, particularly in the experiment with the TodoApp. In essence, this feature reduces the chances that users give up when a bug recurrently appears in the generated code. When this happens, users can readily restore a version that not have this bug. 

\item The code execution and visualization button was also very important for users to quickly check and identify issues in the generated code, including both bugs and layout problems.

\end{itemize}

\subsection{Threats to Validity}
\label{sec::threats_to_validity-NoCodeGPT-evaluation}

The results presented in sections \ref{sec::evaluation_todoapp} and \ref{sec::evaluation_forumapp} are vulnerable to two main threats to validity. 
Firstly, the limited sample size of 14 participants who formulated prompts for the construction of the applications. It is possible that these participants do not represent the entire population of users who intend to use ChatGPT to support end-to-end software construction. However, participants with a similar profile and level of experience to that of P3 from the exploratory study were carefully selected to mitigate this limitation.
Secondly, the systems (ForumApp and TodoApp) defined for construction in the two experiments may not represent the full spectrum of systems that are built from scratch by software developers. However, two well-known applications were selected that follow a common architecture (web-based, with front-end and back-end components) and use popular technologies such as TypeScript, Vue.js, and SQLite, which enhances the generalizability of the findings.

\section{Related Work}
\label{sec::related_work}

In this section, we discuss  recent papers related to our study.  First, we comment on papers related to our central goal of using LLMs to support software construction. After, we discuss papers that rely on LLMs to automate software maintenance tasks, including fixing bugs and writing tests.\\

\noindent{\bf Software Construction:} Researchers from Microsoft Research, GitHub, and MIT conducted a controlled experiment with professional developers who were asked to implement an HTTP server in JavaScript~\cite{peng2023impact}. The researchers report that the treatment group, which had access to GitHub Copilot, was able to complete the proposed task $55.8\%$ faster.  However, we claim that our study uses an application that reflects the practice of software development today. For example, our Q\&A forum includes a front-end (implementing in a widely used framework, Vue.js), an API provided by the back-end (in the form of a set of HTTP end-points), and a SQLite database. 

Le and Zhang evaluated the use of ChatGPT in a very specific context, i.e., to automatically implement parsers for log files~\cite{le2023evaluation}. In a dataset with thousands of log files, the authors report that ChatGPT achieved 71\% accuracy, i.e.,~log messages that were correctly recovered by the parser generated by the tool (from a generic prompt). White and colleagues describe a set of 13 prompt patterns for solving a wide range of software engineering problems, from system design to implementation and maintenance~\cite{white2023chatgpt}. However, these prompts are more abstract and generic than the ones we use in our study. For example, at the requirements level, they verify whether certain requirements were correctly specified or they are proposed to detect ambiguities in requirements specification. On the other hand, our prompts aim to generate usable and runnable software from agile-based requirements (written as user stories). It is also worthnoting that White and colleagues propose six prompt patterns related to code maintenance, evolution, and refactoring tasks, which are outside the scope of our research. Finally, it is also important to mention that in our work, we not only define the prompts but also use them in a real-world context for end-to-end software construction.

Similarly, Sadik and colleagues from the Honda Research Institute comprehensively describe various applications of LLMs in software engineering, including code generation, documentation, bug detection, and refactoring~\cite{sadik2023analysis}. However, the authors' goal is to explore the use of LLMs in software engineering, and as such, they do not apply their prompts in real-world contexts.

Researchers from Meta describe a competitor to GitHub Copilot they are developing internally at the company~\cite{murali2023codecompose}. In addition to describing the design and implementation of the system, called CodeCompose, the authors present results regarding the tool's usage and adoption. They conclude by mentioning that ``in addition to assisting in code implementation (authoring), CodeCompose is introducing other positive effects, such as encouraging developers to generate more documentation, helping them discover new APIs, etc." 

Nguyen and Nadi evaluated the use of GitHub Copilot on a dataset of programming problems (LeetCode)~\cite{msr-copilot}. The correction of the responses generated by Copilot was assessed using the dataset's own test suite. Additionally, the authors evaluated code quality metrics, such as cyclomatic complexity. Similarly to our findings, they conclude that, overall, Copilot's suggestions have low complexity, with no noticeable differences between  programming languages. Mastropaolo and colleagues assessed the robustness of the code generated by GitHub Copilot~\cite{icse2023-bavota}. Their comparison involved two scenarios: code generation from JavaDoc comments and from a semantically modified version of such comments. In a sample of 892 methods from Java projects, the recommendations generated by ChatGPT differed in nearly half of the tests, demonstrating the tool's sensitivity to the prompts provided as input. 

In a recent blog post, Guo explored the advantages and limitations of ChatGPT in complex programming tasks \cite{real_real_world}. As in our study, he highlights the continued need for human guidance and the need of expertise and previous education in Software Engineering for an effective use of the tool.

Dakhel and colleagues examine GitHub Copilot's efficacy in generating solutions for core Computer Science problems~\cite{Dakhel2023}. They compare Copilot's solutions with those produced by human programmers and find that, while Copilot generates solutions for most problems, they exhibit more bugs compared to human solutions. However, the authors identify that the Copilot's buggy solutions are easier to fix than the human ones.

Liu and his colleagues~\cite{liu2023refining} conducted a study that demonstrated ChatGPT's ability to effectively address programming challenges and significantly improve code quality by over 20\%. The study also found that ChatGPT generated functionally correct programs in Python and Java with success rates of 66\% and 69\%, respectively.\\

\noindent{\bf Software Maintenance and Testing:}  Sobani and colleagues evaluated the performance of ChatGPT in bug fixing, using a dataset commonly adopted in this field (QuixBugs)~\cite{SobaniaBHP23}. The authors conclude that the system's performance is considerably better than other approaches proposed in the literature. Specifically, ChatGPT was able to repair 31 out of 40 bugs evaluated in the research. Siddiq and colleagues, on the other hand, report less promising results regarding the use of ChatGPT to implement unit tests ~\cite{siddiq2023exploring}. In fact, in a first dataset, the statement coverage of the automatically generated tests was good (80\%). However, in a second dataset, the results were worse (only 2\% coverage). The authors also report that the generated tests have some smells, such as Duplicate Asserts and Empty Tests. Interestingly, in the three apps constructed in our study, we did not identify any code smells.

Asare, Nagappan, and Asokan evaluate whether the code generated by GitHub Copilot has the same security flaws as code written by developers~\cite{asare2023githubs}. The conclusion was that in 33\% of cases, Copilot essentially replicates the vulnerabilities that exist in a dataset of C and C++ systems.

Pearce and colleagues explore the security implications of code generated by GitHub Copilot~\cite{pearce2021asleep}. The authors evaluate Copilot's performance across three dimensions of code generation, specifically assessing its capabilities in addressing various weaknesses, prompts, and domains. Their analysis covers 1,692 programs generated by Copilot, revealing that approximately 40\% of them exhibit vulnerabilities.

Meta's software engineers conducted a study employing the TestGen-LLM tool~\cite{alshahwan2024automated}. This tool automatically enhances pre-existing human-written tests for Instagram and Facebook platforms using Language Model Models. The results showed that 11.5\% of all test classes extended by TestGen-LLM had improved tests. Moreover, 73\% of the recommendations generated by TestGen-LLM were accepted by Meta's software engineers for implementation in the production environment.

\section{Conclusion}
\label{sec::conclusion}

Language Models are being widely used by Software Engineers. Therefore, in this article, we started by investigating the use of ChatGPT for generating small Web applications by users with limited experience in software development. We concluded that the standard GPT interface was not designed to specifically address this task. For this reason, we proposed, designed, and implemented a new interface for using language models, which provides a more user-friendly experience for building Web apps. This new interface, which we called NoCodeGPT, encapsulates the entire exchange of prompts with the model. As a result, users no longer need to write prompts related to technologies and architecture patterns or copy files to their respective local folders. We also conducted a controlled experiment with 14 students who are beginners in Web development. The results provided convincing evidence that the proposed tool is indeed superior to the standard ChatGPT interface, helping inexperienced developers build small Web apps without writing code.
For example, more than half of the participants (9 out of 14) successfully completed the proposed
applications, while the others completed at least half of the proposed user stories.

As future work, we intend to evaluate the use of our interface with more apps and more developers. We also plan to make it compatible with other language models, including open-source models. Finally, we aim to investigate the construction of new types of applications, such as mobile apps.

\subsection*{Conflict of Interest Statement}

We wish to confirm that there are no known conflicts of interest associated with this publication and there has been no significant financial support for this work that could have influenced its outcome.

\subsection*{Data Availability Statement}

The code of our NoCodeGPT tool is available at:
\href{https://github.com/mauricioms/nocodegpt}{https://github.com/mauricioms/nocodegpt}. This repository also contains the prompts from the first (exploratory) study and the evaluation of NoCodeGPT.

% ... 
%
%


\subsection*{ORCID}

Maurício Monteiro: \orcidlink{0009-0005-2878-7962} \url{https://orcid.org/0009-0005-2878-7962}

\subsection*{Acknowledgments}

Our research is supported by grants from CNPq and FAPEMIG.

\small
\bibliographystyle{plain}
\bibliography{bib}

\begin{thebibliography}{10}

\bibitem{alshahwan2024automated}
Nadia Alshahwan, Jubin Chheda, Anastasia Finegenova, Beliz Gokkaya, Mark Harman, Inna Harper, Alexandru Marginean, Shubho Sengupta, and Eddy Wang.
\newblock Automated unit test improvement using large language models at meta.
\newblock {\em CoRR}, abs/2402.09171, 2024.

\bibitem{asare2023githubs}
Owura Asare, Meiyappan Nagappan, and N.~Asokan.
\newblock Is {GitHub's Copilot} as bad as humans at introducing vulnerabilities in code?
\newblock {\em Empirirical Software Engineering}, 28(6):129, 2023.

\bibitem{geng2023large}
Mingyang Geng, Shangwen Wang, Dezun Dong, Haotian Wang, Ge~Li, Zhi Jin, Xiaoguang Mao, and Xiangke Liao.
\newblock Large language models are few-shot summarizers: Multi-intent comment generation via in-context learning, 2024.

\bibitem{real_real_world}
Philip {Guo}.
\newblock Real-real-world programming with {ChatGPT}, 2023.
\newblock Available online at: \url{https://www.oreilly.com/radar/real-real-world-programming-with-chatgpt/}, last accessed on Oct 2023.

\bibitem{le2023evaluation}
Van{-}Hoang Le and Hongyu Zhang.
\newblock An evaluation of log parsing with {ChatGPT}.
\newblock {\em CoRR}, abs/2306.01590, 2023.

\bibitem{liu2023refining}
Yue Liu, Thanh Le{-}Cong, Ratnadira Widyasari, Chakkrit Tantithamthavorn, Li~Li, Xuan{-}Bach~Dinh Le, and David Lo.
\newblock Refining {ChatGPT}-generated code: Characterizing and mitigating code quality issues.
\newblock {\em CoRR}, abs/2307.12596, 2023.

\bibitem{icse2023-bavota}
Antonio Mastropaolo, Luca Pascarella, Emanuela Guglielmi, Matteo Ciniselli, Simone Scalabrino, Rocco Oliveto, and Gabriele Bavota.
\newblock On the robustness of code generation techniques: An empirical study on {GitHub Copilot}.
\newblock In {\em 45th {IEEE/ACM} International Conference on Software Engineering}, pages 2149--2160, 2023.

\bibitem{Dakhel2023}
Arghavan {Moradi Dakhel}, Vahid Majdinasab, Amin Nikanjam, Foutse Khomh, Michel~C. Desmarais, and Zhen Ming~(Jack) Jiang.
\newblock Github copilot ai pair programmer: Asset or liability?
\newblock {\em Journal of Systems and Software}, 203:111734, 2023.

\bibitem{murali2023codecompose}
Vijayaraghavan Murali, Chandra~Shekhar Maddila, Imad Ahmad, Michael Bolin, Daniel Cheng, Negar Ghorbani, Renuka Fernandez, and Nachiappan Nagappan.
\newblock {CodeCompose}: {A} large-scale industrial deployment of {AI}-assisted code authoring.
\newblock {\em CoRR}, abs/2305.12050, 2023.

\bibitem{msr-copilot}
Nhan Nguyen and Sarah Nadi.
\newblock An empirical evaluation of {GitHub Copilot}'s code suggestions.
\newblock In {\em 19th International Conference on Mining Software Repositories (MSR)}, pages 1--5, 2022.

\bibitem{pearce2021asleep}
Hammond Pearce, Baleegh Ahmad, Benjamin Tan, Brendan Dolan{-}Gavitt, and Ramesh Karri.
\newblock Asleep at the keyboard? assessing the security of {GitHub Copilot}'s code contributions.
\newblock In {\em 43rd {IEEE} Symposium on Security and Privacy}, pages 754--768, 2022.

\bibitem{peng2023impact}
Sida Peng, Eirini Kalliamvakou, Peter Cihon, and Mert Demirer.
\newblock The impact of {AI} on developer productivity: Evidence from {GitHub Copilot}.
\newblock {\em CoRR}, abs/2302.06590, 2023.

\bibitem{sadik2023analysis}
Ahmed~R. Sadik, Antonello Ceravola, Frank Joublin, and Jibesh Patra.
\newblock Analysis of {ChatGPT} on source code.
\newblock {\em CoRR}, abs/2306.00597, 2023.

\bibitem{shin2023prompt}
Jiho Shin, Clark Tang, Tahmineh Mohati, Maleknaz Nayebi, Song Wang, and Hadi Hemmati.
\newblock Prompt engineering or fine tuning: An empirical assessment of large language models in automated software engineering tasks, 2023.

\bibitem{siddiq2023exploring}
Mohammed~Latif Siddiq, Joanna C.~S. Santos, Ridwanul~Hasan Tanvir, Noshin Ulfat, Fahmid~Al Rifat, and Vinicius~Carvalho Lopes.
\newblock Exploring the effectiveness of large language models in generating unit tests.
\newblock {\em CoRR}, abs/2305.00418, 2023.

\bibitem{SobaniaBHP23}
Dominik Sobania, Martin Briesch, Carol Hanna, and Justyna Petke.
\newblock An analysis of the automatic bug fixing performance of {ChatGPT}.
\newblock In {\em {IEEE/ACM} International Workshop on Automated Program Repair (APR@ICSE 2023)}, pages 23--30. {IEEE}, 2023.

\bibitem{white2023chatgpt}
Jules White, Sam Hays, Quchen Fu, Jesse Spencer-Smith, and Douglas~C. Schmidt.
\newblock {ChatGPT} prompt patterns for improving code quality, refactoring, requirements elicitation, and software design, 2023.

\bibitem{XiaWZ23}
Chunqiu~Steven Xia, Yuxiang Wei, and Lingming Zhang.
\newblock Automated program repair in the era of large pre-trained language models.
\newblock In {\em 45th {IEEE/ACM} International Conference on Software Engineering}, pages 1482--1494, 2023.

\end{thebibliography}

\end{document}